\documentclass[11pt]{article}
\usepackage{library}
\usepackage[dvipsnames]{xcolor}
\usepackage{stmaryrd}
\usepackage{gitinfo}
\usepackage[ruled,vlined,linesnumbered]{algorithm2e}

\newcommand{\Res}{\mathrm{Res}}

\newcommand{\Esym}{\mathsf{Esym}}

\newcommand{\inbrace}[1]{\left\{ #1 \right\}}
\newcommand{\indsquare}[1]{\llbracket #1 \rrbracket}

\newcommand{\N}{\mathbb N}

\newcommand{\K}{\mathbb K}
\newcommand{\F}{\mathbb F}
\newcommand{\Q}{\mathbb Q}

   \newcommand{\cB}{{\mathcal{B}}}
\newcommand{\cC}{{\mathcal{C}}}

   \newcommand{\cT}{{\mathcal{T}}}

\usepackage{amsthm}
\usepackage{aliascnt}

\usepackage[colorlinks]{hyperref}
\usepackage[nameinlink,capitalise]{cleveref}

\hypersetup{
  linkcolor=[rgb]{0,0,0.4},
  citecolor=[rgb]{0, 0.4, 0},
  urlcolor=[rgb]{0.6, 0, 0}
}

\newtheorem{theorem}{Theorem}[section]

\newaliascnt{lemma}{theorem}
\newtheorem{lemma}[lemma]{Lemma}
\aliascntresetthe{lemma}

\newaliascnt{proposition}{theorem}

\aliascntresetthe{proposition}

\newaliascnt{corollary}{theorem}
\newtheorem{corollary}[corollary]{Corollary}
\aliascntresetthe{corollary}

\theoremstyle{definition}

\newaliascnt{definition}{theorem}
\newtheorem{definition}[definition]{Definition}
\aliascntresetthe{definition}

\AtEndEnvironment{definition}{\hfill$\square$}

\newaliascnt{remark}{theorem}
\newtheorem{remark}[remark]{Remark}
\aliascntresetthe{remark}

\newaliascnt{example}{theorem}

\aliascntresetthe{example}

\newaliascnt{claim}{theorem}

\aliascntresetthe{claim}

\newtheorem*{proofsketch}{Proof Sketch}
\AtEndEnvironment{proofsketch}{\hfill$\square$}

\crefname{theorem}{Theorem}{Theorems}
\crefname{lemma}{Lemma}{Lemmas}
\crefname{proposition}{Proposition}{Propositions}
\crefname{corollary}{Corollary}{Corollaries}
\crefname{definition}{Definition}{Definitions}
\crefname{remark}{Remark}{Remarks}
\crefname{example}{Example}{Examples}
\crefname{claim}{Claim}{Claims}

\numberwithin{equation}{section}

\newcommand{\X}{\mathbf{X}}
\def\poly{\operatorname{poly}}

\SetKwInput{KwOracle}{Oracle access}
\DontPrintSemicolon

\newif\ifdraft
\drafttrue          

\newcommand{\myignore}[1]{}
\newcommand{\gitinfonote}{git info:~\gitAbbrevHash\;,\;(\gitAuthorIsoDate)\; \;\gitVtag}

\newcounter{todo}
\usepackage[T1]{fontenc}

\newcommand{\gitinfonotecolour}{white}
\newcommand{\easteregg}{The proof is whitebox, not the AI.}

\title{Deterministic factorization of constant depth circuits from \emph{whitebox} PIT}
\author{
     {Somnath Bhattacharjee \thanks{University of Toronto, Canada. Email: \texttt{somnath.bhattacharjee@mail.utoronto.ca}. Research partially supported by an NSERC Discovery Grant}}
     \and
     {Mrinal Kumar \thanks{Tata Institute of Fundamental Research, Mumbai, India. Email: \texttt{\{mrinal, shanthanu.rai, ramprasad\}@tifr.res.in}.  Research supported by the Department of Atomic Energy, Government of India, under project number RTI400112, and in part by research grants from Google, SERB and Premji Invest. }}
          \and
     {Shanthanu S. Rai\footnotemark[2]}
     \and
     {Varun Ramanathan \thanks{University of Copenhagen, Denmark. Email: \texttt{vara@di.ku.dk}. Supported by the European Research Council (ERC) under grant agreement no. 101125652 (ALBA).}}
     \and
     {Ramprasad Saptharishi\footnotemark[2]}
     \and
     {Shubhangi Saraf\thanks{University of Toronto, Canada.
     Email: \texttt{shubhangi.saraf@utoronto.ca}.
     Research partially supported by the McLean Award
     and an NSERC Discovery Grant.}}
}
\date{}

\begin{document}
\maketitle
\begin{abstract}
We show that over the field of rational numbers, there is a deterministic algorithm for factorization of constant depth circuits that makes calls to a (whitebox or blackbox) PIT algorithm for such circuits. The time complexity of the algorithm (and hence the number of calls to the PIT oracle) is polynomially bounded in the size and degree of the input circuit. Our reduction to PIT is fairly simple and general and applies as it is to other natural classes of algebraic circuits like formulas and algebraic branching programs. 

Prior to this work, the known deterministic factorization algorithms for constant depth circuits crucially relied on \emph{blackbox} PIT [Bhattacharjee et al.]. Our main technical ingredient is an efficient, deterministic algorithm that, when given access to sufficiently high degree truncations of \emph{all} the power series roots of a constant depth circuit, uses oracle calls to any (whitebox or blackbox) PIT algorithm for these circuits, and partitions the roots into the sets of conjugates; thereby identifying the irreducible factors. 

For our proof, we rely on a root recombination criterion using logarithmic derivatives [Sasaki et al., Bostan et al., Lecerf] 
and show that this criterion can be efficiently and deterministically implemented for power series roots of constant depth circuits using access to any PIT oracle.  On the way, we discuss an algorithm  that may be of independent interest: it takes as input a matrix whose entries are multivariate constant depth circuits over a field $\F$, requires oracle access to any PIT algorithm for these circuits and outputs a basis for the $\F$-kernel of this matrix in polynomial time. 

\end{abstract}

\section{Introduction}

The complexity of polynomial factorization as an algorithmic task and the relationship between the circuit complexities of a polynomial and that of its factors are questions of fundamental interest in computational algebra and algebraic complexity. Both these families of questions saw substantial advances in the 1980s in a sequence of results of Kaltofen (e.g. see \cite{Kaltofen89, KT90, vzGK85, Burgisser2004Factors}) who showed both a \emph{closure} result for algebraic circuits under factorization, and gave efficient randomized algorithms for this task (see, for instance, the survey of Forbes and Shpilka \cite{FS15} for a brief discussion of the literature). More formally, the closure result showed that if an $n$ variate degree $d$ polynomial $f$ over the field of rational numbers\footnote{Slightly weaker results of a similar flavor also hold over finite fields.} is computable by an algebraic circuit of size $s$, then all its irreducible factors can be computed by circuits of size $\poly(s,d)$. The algorithmic results consisted of efficient randomized algorithms that compute these irreducible factors in randomized $\poly(s,d)$ time, when given the circuit for $f$ as input.  Remarkably, these algorithms work in both the \emph{blackbox} and the \emph{whitebox} settings: in the blackbox setting the algorithm is just given evaluation access to a blackbox that contains $f$ and it outputs blackboxes corresponding to the irreducible factors of $f$, and in the whitebox setting the algorithm has access to the internal wirings of a circuit for $f$, and it outputs explicitly, the circuits for its irreducible factors.

Subsequent to these results, perhaps the two most important research directions in this area have been that of extending the closure result to the natural structured subclasses of algebraic circuits, e.g formulas/branching programs/constant depth circuits, and that of designing efficient deterministic algorithms for polynomial factorization. There has been substantial progress on the question of closure results in recent years, and we now know that algebraic branching programs \cite{SinhababuThierauf}, formulas \& constant depth circuits \cite{BKRRSS25a} are all \emph{closed} under factorization, and so is the complexity class VNP \cite{ChouKS2019}. 

The progress on the question of designing deterministic factorization algorithms has been comparatively subdued. A work of Kopparty, Saraf and Shpilka \cite{KSS15} showed that derandomizing polynomial identity testing (PIT) is both necessary and sufficient for derandomizing polynomial factorization for general algebraic circuits. While the reduction from PIT to factorization is relatively simple, the other direction required a clean algebraic understanding of the necessity and use of randomness in Kaltofen's factorization algorithms. This equivalence is also variant preserving in the sense that blackbox PIT for general circuits is equivalent to blackbox polynomial factorization, and whitebox PIT is equivalent to factorization in the whitebox model. 

An intriguing aspect of the equivalence result of Kopparty et al. is that it does not appear to \emph{scale down}: while it is easy to see that PIT reduces to factorization for models like formulas and constant depth circuits, it was not clear if derandomizing PIT for these models would imply efficient deterministic factorization algorithms for them. This research direction was explored in a sequence of recent results \cite{KRS24, KRSV26, BKRSS25}, and led to the results of Bhattacharjee et al. \cite{BKRSS25} that gave a subexponential time deterministic algorithm for factorization of constant depth algebraic circuits. In a follow up work \cite{BKRRSS25a}, Bhattacharjee et al.  established closure results for constant depth circuits (and other classes). They also used this closure result to simplify many of the technical details of the algorithm in \cite{BKRSS25}, and obtained a clean statement connecting polynomial factorization to deterministic blackbox PIT for constant depth circuits.  When combined with the known deterministic subexponential time algorithms for blackbox PIT for constant depth circuits,\footnote{That were proved via a combination of superpolynomial lower bounds for these circuits from \cite{LST2025journal} with some partial closure results for such circuits in \cite{ChouKS2019}.} this yielded another deterministic, subexponential time algorithm for factorization of these circuits.  

Very curiously, the deterministic factorization algorithms in \cite{BKRSS25} and \cite{BKRRSS25a} crucially relied on the fact that the known subexponential time PIT algorithms for constant depth circuits were in the \emph{blackbox} model. In particular, it was not known if a similar deterministic factorization algorithm can be obtained from a whitebox PIT algorithm for constant depth circuits. In this work, we answer this question in affirmative. We state the theorem informally here, and refer to \autoref{thm: Whitebox PIT implies Factorization} for a formal statement. 

\begin{theorem}[Informal]\label{thm:intro-main}
    There is a deterministic algorithm that takes as input a size $s$, degree $d$ multivariate algebraic circuit of constant depth  over the rational numbers,\footnote{Throughout this paper, we work over the field of rational numbers.  Assuming the correctness of the recent efficient deterministic univariate factorization algorithms over finite fields \cite{openAI}, the results also extend to finite fields of sufficiently large characteristic.} requires oracle access to a PIT algorithm for constant depth circuits in either blackbox or whitebox setting, and outputs constant depth circuits for the irreducible factors of the input polynomial, together with their multiplicities, in $\poly(s,d)$ time. 

    Furthermore, a similar reduction from factorization to deterministic PIT holds for algebraic formulas as well. 
\end{theorem}

For algebraic branching programs a similar reduction can be seen in the work of Sinhababu \& Thierauf \cite{SinhababuThierauf}. 
Thus,  \autoref{thm:intro-main}, together with the reduction in \cite{BKRRSS25a} give a scaled down version of the equivalence between PIT and polynomial factorization for general circuits shown by Kopparty, Saraf and Shpilka \cite{KSS15}. 

As of now, we do not have deterministic PIT algorithms for constant depth algebraic circuits in the whitebox setting that are faster than those in the blackbox setting. However, it is typically the case in derandomization literature that whitebox algorithms are often discovered before blackbox PIT algorithms for many classes of algebraic circuits and they tend to be faster (see \cite[Sections~4.6 and~4.8]{SY10} for a general overview). 
\autoref{thm:intro-main} shows that any improvement in the state of art for PIT for constant depth circuits, even in the whitebox setting would result in an improvement in the state of art for polynomial factorization for these circuits.  We remark that the various structural results for constant depth circuits in the work of Andrews \& Wigderson \cite{AW24} and those in \cite{ BKRRSS25a, BKRRSS25b} play a crucial role in the proof of \autoref{thm:intro-main}. However, given these structural results, the final factorization algorithm for constant depth circuits obtained in this work appears to be technically simpler in many of its details than those known from prior work \cite{BKRSS25, BKRRSS25a}. 

On the way to the proof of \autoref{thm:intro-main}, we show the following result that may be of independent interest. 

\begin{theorem}[Kernel basis computation over base field]\label{thm:intro-kernel-basis}
    There is a deterministic algorithm that takes as input a matrix $M(\X)$ of dimension $m$ whose entries are multivariate polynomials computable by size $s$, degree $d$ algebraic circuits of constant depth over a field $\F$ (satisfying $|\F|>d$), requires oracle access to a PIT algorithm for constant depth circuits over $\F$ in either blackbox or whitebox setting, and outputs a basis for the $\F$-kernel of $M(\X)$ in time $\poly(s,m,d)$. 

    Here, by an $\F$-kernel, we mean the $\F$-linear space of  $m$-dimensional vectors over $\F$ that are in the kernel of $M(\X)$. 
\end{theorem}
The proof of \autoref{thm:intro-kernel-basis} was discovered with AI assistance (ChatGPT Astra).
{\let\thefootnote\relax
\footnotetext{\textcolor{\gitinfonotecolour}{\gitinfonote \easteregg}
}}

Yet another independently interesting application of \autoref{thm:intro-kernel-basis} is to the question of linear independence testing of polynomials. In particular, we have the following immediate corollary. 
\begin{corollary}\label{cor:lin-ind-test}
 There is a deterministic algorithm that takes as input polynomials $f_1, f_2, \dots, f_m$ given by circuits of size at most $s$ and degree $d$, requires oracle access to a (whitebox or blackbox) PIT algorithm for circuits of the form $\left(\sum_{i}\alpha_i f_i \right)$ for $\alpha_i \in \F$, and decides if $f_1, \dots, f_m$ are linearly independent over the field $\F$ in time $\poly(s,d, m)$. 
\end{corollary}
While this problem has been studied in the literature (see, for instance, \cite{Kay10, CKW11, BhargavaSV2026}), it is unclear if this specific reduction to \emph{whitebox} PIT was known to exist. The corollary can be instantiated with additional guarantees on the structure of the circuits for $f_1, f_2, \ldots, f_m$ to obtain deterministic algorithms for testing the linear independence of polynomials in interesting settings. For instance, if the input polynomials are all sparse, then using the known deterministic PIT for sparse polynomials, this can be done in polynomial time even in the blackbox setting. Similarly, we can obtain a polynomial time deterministic algorithm for testing linear independence of ROABPs (in the same order) in the whitebox setting using the PIT algorithm of Raz \& Shpilka \cite{RazShpilka}. While neither of these two applications appear to be new, the algorithms obtained using Corollary \autoref{cor:lin-ind-test} appear to be much more modular  and essentially remain the same at a high level, even as the circuit classes containing the input change. 

\subsection{An overview of the proofs}
\label{sec:proof-overview}

Let $f(\X,Y)$ be the input polynomial given as a small constant depth circuit with rational coefficients. Then, the main steps in the factorization algorithm on input $f$ in various proofs in the literature, and in particular in \cite{BKRSS25, BKRRSS25a} are as follows. 
\begin{enumerate}
    \item {\textbf{Preprocessing: }} In this step, by an invertible linear transformation of the form $X_i \to X_i + a_iY$, for $a_i\in \F$, it is ensured that $f$ is monic in $Y$. At this point, using standard techniques, $f$ can be assumed to be square free. Now, by shifting the $\X$-variables further if necessary, we reduce to the setting that $f(\mathbf{0},Y)$ is square free and splits into distinct linear factors\footnote{We get rid of this assumption in \cref{sec:logarithmic-derivative-method} by using Hensel Lifting.} over $\F$. We now pick a new formal variable $T$, and substitute every variable $X_i$ by $X_iT$. The new polynomial, which we refer to as $F(T,Y)$ here is viewed as a bivariate in $T, Y$, with its coefficients in $\F(\X)$. 

    \item {\textbf{Computing power series roots: }} Using the squarefreeness of $f(\mathbf{0},Y)$, and hence $F(0,Y)$, we get from \cref{thm:truncations-of-power-series-roots-of-constant-depth-circuits} that the $Y$-roots of $F$ are power series in $T$, with coefficients in $\F(\X)$. Furthermore, sufficiently high degree truncations of these roots are computable by small constant depth circuits, and these circuits can be efficiently and deterministically constructed from a circuit for $F$ and the roots of $F(0,Y)$. 

    \item \textbf{Obtaining irreducible factors: } The final step of the factoring algorithms is to identify the power series roots of $F$ that come from the same irreducible factor of $F$. Once this partition of the set of roots is obtained, the irreducible factors of $F$ (and hence $f$) can be constructed efficiently, using 
    \[
    f_i(T,Y) = \prod_{i \in S} (Y - \phi_i(T)) \bmod T^{\deg_T(f) + 1} ,  
    \]
    where, $S \subseteq [\deg_Y(F)]$ is the set of indices of all power series roots of the irreducible factor $f_i$. 
\end{enumerate}
The power series root computation step above is completely deterministic, and randomness is used only for preprocessing and for the root combination steps. Of these, it is not difficult to see (and is discussed in detail in the proofs in \cite{KSS15,BKRSS25,BKRRSS25a}) that randomness for preprocessing can be removed with access to PIT algorithms. For the setting of constant depth circuits, we need the additional amazing fact proved by Andrews \& Wigderson \cite{AW24} (also see \cite{BKRRSS25b}) that Resultants and Discriminants are computable by constant depth circuits. This allows us to show that for constant depth circuits, the preprocessing phase of the factorization algorithms  can be derandomized using an access to a PIT oracle, in either the blackbox or the whitebox setting. A crucial (and simple) part of this argument is that one can efficiently and deterministically solve the search version of PIT, where the goal is to find a point where the input polynomial evaluates to a non-zero value, when given  access to a PIT oracle (see \cref{lem:decision-to-search-PIT}). 

The main technical difficulty that needed to be addressed for the proofs  in this paper, and in \cite{BKRSS25, BKRRSS25a} was to derandomize the root combination step to identify roots that correspond to each irreducible factor of $F$. More concretely, we have the following problem: we have constant depth circuit computing a polynomial $F(T,Y) \in \F(\X)[T,Y]$ with $\deg_Y(F) \leq d, \deg_T(F) \leq D$, and also have constant depth circuits for polynomials $B_1, B_2, \ldots, B_d \in \F(\X)[T,Y]$ such that for every $i$, 
\[
F(T, B_i) \equiv 0 \bmod T^m, 
\]
for some $m = \poly(d,D)$ that is large enough. Our goal is to partition the $B_i's$ into sets corresponding to the irreducible factors of $F$. In \cite{BKRSS25, BKRRSS25a}, this was done by using blackbox PIT for constant depth circuits (given via a hitting set generator): the hitting set is used to set the $\X$ variables to field constants in $\F$. This gives us many bivariate polynomials $g(T,Y)$ in $T, Y$ with coefficients over the base field $\F$, that can be factorized deterministically in polynomial time using the algorithm outlined at the beginning of this section.  The proofs in \cite{BKRSS25, BKRRSS25a} proceed by using the properties of the hitting set generator to guarantee that the correct partition of the power series roots of most of the bivariates $g$ into buckets corresponding to their irreducible factors is in one-one correspondence with the correct partitioning of the power series roots of $F(T,Y) \in \F(\X)[T,Y]$. This part of the proof uses the crucial fact that a hitting set generator for constant depth circuits will maintain the non-zeroness of an exponentially large set of constant depth circuits at once. This argument is only used in the proof of correctness of the algorithm, and is not implemented in the algorithm itself. Thus, this argument crucially relies on the fact that we have access to a hitting set generator (and not just a whitebox algorithm) for PIT of constant depth circuits. 

For our proof, we give an alternative way to correctly combine the power series roots into irreducible factors. 
The proof that we present in \cref{sec:logarithmic-derivative-method} is based on the so-called \emph{logarithmic derivative} idea that is very well studied in the factorization literature (see \cite{Sasaki1991, Sasaki1993, BLSSW04, Lecerf2006, Lec06, Lecerf2010} and the references therein), albeit in the bivariate or the dense representation setting. For this overview, we try to provide some intuition into the method. 

Since $F(0,Y)$ is square free, all of its roots, that we assume are present in the base field $\F$, must be distinct. 
Since $B_1, \dots, B_d$ are power series roots of $F(T,Y)$ obtained by \emph{lifting} the roots of $F(0,Y)$, they must have distinct constant terms.  Suppose $H(T,Y)$ is an irreducible factor of $F(T,Y)$. Then, there is a subset $S \subseteq [d]$, with $|S| = \deg_Y(H)$ such that $\{B_i : i \in S\}$ is the set of truncations of all power series $Y$-roots of $H$. 

Since $H$ is a factor of $F$, we have that its $T$-degree is at most $D$. Thus, the elementary symmetric polynomials of the power series roots of $H$ must be polynomials in $\F(\X)[T]$ and must have $T$-degree at most $D$. Thus, in the expressions for the elementary symmetric polynomials of the power series roots of $H$, there must be significant cancellations, leading to these expressions yielding a low degree polynomials in $T$ on simplification.  The root recombination method is based on the fact that the converse of this observation is also essentially true: if all elementary symmetric polynomials of a subset $\hat{A}$ of power series roots of $F$ are low degree polynomials in $T$, then $\hat{A}$ corresponds to a (not necessarily irreducible) factor of $F$. 

While mathematically insightful, it is not immediately clear if we can \emph{efficiently} search for  subsets $\hat{A}$ of the power series roots of $F$ that have this cancellation property, since trying all subsets is not within the budget. This approach seems more amenable to efficient algorithms if we work with power sum symmetric polynomials instead. Due to Newton-Girard identities that let us express power sums as bounded degree polynomials of elementary symmetric polynomials (and vice-versa), and the above discussion, we get that for any subset $\hat{A}$ of the power series roots of $F$, the power sum symmetric polynomials of degree\footnote{Recall that unlike the elementary symmetric polynomials that are zero when the degree exceeds the number of variables, power sum polynomials do not vanish at high degrees.} at most $d$  on $\hat{A}$ are all low degree polynomials if and only if $\hat{A}$ corresponds to a factor of $F$. The advantage is that the search for subsets $\hat{A}$ now looks a little more tractable: consider formal variables $u_1, u_2, \dots, u_d$ and  the linear constraints 
\[
\forall \ell \leq d, \quad \quad  \deg_T\left(\sum_i u_i \phi_i^{\ell}\right) \leq  \hat{D}_\ell , 
\]
where $\phi_i$'s are the true (untruncated) power series roots of $F$, and each $\hat{D}_l \in \N$ is an appropriately chosen parameter. Once we have such a bound for $\ell \leq d$, a similar ($\ell$-dependent) bound follows for \emph{all} $\ell$ via the recurrence $p_\ell=-a_1 p_{\ell-1}-\cdots-a_d p_{\ell-d}$ where each $p_\ell$ denotes $\sum_{i
\in [d]} {u_i \phi_i^\ell}$, and $F = Y^d + a_1 Y^{d-1} + \dots + a_d$. A succinct equivalent representation of this system obtained by tying together the powers of $\phi_i$ into a generating function $(1/(Y-\phi_i))$ is as follows: 
\[
\deg_T\left([Y^{-\ell-1}]\sum_i u_i \cdot \frac{1}{(Y-\phi_i(T))}\right) \leq \hat{D}_\ell. 
\]
Each $1/(Y-\phi_i(T))$ is the ``logarithmic derivative'' of $Y-\phi_i(T)$ and similarly, $\sum_{i \in [d]}\frac{1}{Y-\phi_i(T)}$ is the logarithmic derivative of $F$; see \cite{LogDerivative} for a general definition. Clearing the denominator with $F(T,Y) = \prod_{i\in [d]}(Y-\phi_i(T))$ give us the system:
\[
\deg_T\left([Y^{-\ell-1}]\left(\sum_i u_i \cdot \prod_{j \neq i}(Y-\phi_j(T))\right) \right) \leq \tilde{D}_\ell  
\]
for some appropriately scaled $\tilde{D}_\ell$. But this is a system that uses power-series roots. Instead, we work with the following very closely related linear system  described in \cite{Sasaki1991, Sasaki1993, BLSSW04, Lecerf2006, Lec06, Lecerf2010}:
\[
\deg_T\left( \left(\sum_i u_i \cdot \prod_{j \neq i}(Y-B_j(T))\right) \bmod T^m\right) \leq {D}.  
\]
Note that this only relies on the truncations of the power series roots, and $\tilde{D}_\ell$ are all set to $D$. 

This is clearly a homogeneous linear system in $u_1, \dots, u_d$. If $F_1, F_2, \dots, F_r$ are all the irreducible factors of $F$ and $w_1, w_2, \ldots, w_r \in \{0,1\}^d$ are vectors defined by setting the $j$th coordinate of $w_i$ to $1$ if and only if $B_j$ is a truncated power series root of $F_i$, then we observe that $w_1, w_2, \dots, w_r$ are indeed solutions of this linear system: setting the $u$-variables according to $w_i$ reduces the expression $$\sum_i u_i \cdot \prod_{j \neq i}(Y-B_j(T)) \bmod T^m$$ to $\partial_Y(F_i)\frac{F}{F_i}$, and for $m \gg D$, this clearly has $T$-degree at most $D$.  

A key technical observation for the proof  is that $w_1, w_2, \dots, w_r$ are in-fact a basis of the $\F$-linear space of solutions of this linear system in $\F^d$! Furthermore, they are the unique basis of this solution space (in $\F^d$) in the \emph{reduced row echelon} form! Thus, if we could somehow find the unique basis of the solution space in the reduced row echelon form, we would be done. The challenge is to do this deterministically, perhaps using a PIT oracle, assuming that $F$ has a small constant depth circuit. 

From the results in \cite{BKRRSS25a}, we know that each $B_j$ has a small constant depth circuit. This implies that the linear system in $u_1, \ldots, u_d$ reduces to finding a vector in $\F^d$ in the kernel of a matrix $M$. The entries of the matrix $M$ are $n$-variate polynomials in $\F[\X]$ that are computable by small constant depth circuits. An algorithm for this problem, using access to a whitebox PIT oracle for constant depth circuit is the final technical ingredient of our proof. Several variants of polynomial factorization algorithms in literature (e.g \cite{BKRRSS25a, BKRRSS25a}) involve a linear system solving step. However, these systems are typically over the field $\F(\X)$. Our algorithm for finding a basis for the $\F$-kernel of $M$ crucially relies on the fact that we are looking for an $\F$-kernel and not an $\F(\X)$-kernel (whose computation we believe is harder to derandomize). This algorithm, presented in \cref{sec:logarithmic-derivative-method} was discovered with AI assistance and happens to be surprisingly clean and elegant in hindsight. Indeed we are optimistic that such an algorithm might be of independent interest. Given such a basis for the kernel of $M$, we compute the unique basis of this space in the reduced row echelon form, and this gives us $w_1, \dots, w_r$, thereby completing the recombination step.

\subsection*{AI disclosure } The proof of \autoref{thm:intro-kernel-basis} was discovered using ChatGPT Astra. The rest of the paper including all the technical ideas and the writing are due to its human authors. 

\section{Preliminaries}
\subsection{Notation}
\begin{itemize}
\item Throughout this paper we will be working over a characteristic zero field (such as $\Q$). 
\item $[n]$ denotes the set $\{1, 2, \dots, n\}$.
\item For any vector $v\in\F^n$, support of $v$ is the set $\{i \in [n] : v_i \neq 0 \}$.
\item We will use $\X$ to represent the set of variables $\{X_1,\dots,X_n\}$. We use $\F[T]$ and $\F\indsquare{T}$ to refer to the ring of polynomials and formal power series in $T$ respectively. 
\item For any power series or polynomial $P(Z)$, $[Z^k]\{P\}$ will denote the coefficient of $Z^k$ in $P(Z)$.
\item For a polynomial $P(\X) \in \F[\X]$, $\deg(P)$ denotes the total degree of $P$, and for an $X_i \in \X$, $\deg_{X_i}(P)$ denotes the degree of $P$ in $X_i$.
\item Let $\{v_1,\dots,v_r\}$ be a set of vectors over a field $\mathbb{K}$. Then for any subfield $\F\subseteq\K$, we denote $\operatorname{span}_\F(v_1,\dots,v_r)$ to be the space of $\F$-linear combinations of $v_i$s. 
\item Let $M \in \F^{m \times n}$ be a $m \times n$ matrix. Define $\ker_\F(M)$ to denote the kernel of $M$.
\item For any polynomial (or formal power series) $P(T)$, $P(T) \bmod T^{D+1}$ will denote the polynomial obtained by discarding all terms of degree at least $D+1$.
\item We will use $\Esym_r\{v_1,\dots,v_m\}$ to denote the $r$-th elementary symmetric polynomial of the set $\{v_1,\dots,v_m\}$, i.e. $\Esym_r\{v_1,\dots,v_m\}=\sum_{S\subseteq[m],|S|=r}\prod_{i\in S}v_i$.
\end{itemize}

\subsection{Algebraic Circuits}

We will define the standard model of computation, namely algebraic circuits, and record some useful and well-known facts pertaining them which will be useful later own.

\paragraph{Algebraic Circuits:} An algebraic circuit $C$ is a DAG with a designated output node, wherein the leaf nodes are labeled by variables and field constants and the internal nodes are labeled by $+$ and $\times$. The circuit $C$ computes the polynomial $P(\X)$ at the output gate, by inductively computing it at each node from leaf nodes to the output node. The size of the circuit $C$ is the number of wires in $C$, and the depth is the length of the longest path from the leaf node to the output node.

Michael Ben-Or observed that algebraic circuits are ``closed'' under polynomial interpolation. This means given a circuit $C$ for a polynomial $f(\X, Y)$, there exist circuits for coefficients of $f(\X, Y)$ (in $Y$) which are not much larger in size than $C$. Moreover, there is an efficient algorithm to compute these circuits from $C$.

\begin{lemma}[Interpolation]
    \label{lem:interpolation} Let $\F$ be a sufficiently large field and $f(\X,Y)\in\F[\X,Y]$ be a $(n+1)$-variate polynomial with $\deg_Y(f)=d$. Suppose $f(\X, Y)$ is computable by a circuit of size $s$ and depth $\Delta$. Then, there is an algorithm that outputs a multi-output circuit of size $\poly(s,n,d)$ and depth $\Delta+O(1)$ and runs in time $\poly(n, s, d)$. The multi-output circuit outputs $[Y^m]f(\X,Y)$ for all $m\in\{0,1,\dots,d\}$.
\end{lemma}

\paragraph{White-box PIT:} One of the fundamental problems in algebraic complexity is the Polynomial Identity Testing (PIT) problem: Given a circuit $C$ (or any well-defined algebraic model), does it compute the $0$ polynomial? If the internal wiring in the circuits are revealed to us, then this is the question of PIT in the white-box model (succinctly referred to as the white-box PIT problem). Coming up with efficient PIT algorithms has important connections to lower bounds in different algebraic models, polynomial factorization, etc \cite{KI04,SY10,KSS15,KS19-hardness-randomness}. Using an algorithm for the decision version of the problem, that is determining if a circuit $C$ (say with $n$ inputs) is non-zero, one can find a point $\alpha \in \F^n$ such that $C$ is non-zero at $\alpha$ (namely, the search version).

\begin{lemma}[Decision to Search PIT]
    \label{lem:decision-to-search-PIT} Let $\F$ be a field such that $|\F| > d$ and $f(\X)\in\F[\X]$ be an $n$-variate, degree $d$ non-zero polynomial from a circuit class $\cC$ . Then there is an $\poly(n,d)$ algorithm to compute an $\mathbf{\alpha}\in\F^n$ such that $f(\mathbf{\alpha})\neq0$ that uses oracle queries to a white-box PIT algorithm for $\cC$.
\end{lemma}

\subsection{Resultant and the Poisson Formula}
\label{sec:resultant}

The resultant of $f(Y)$ and $g(Y)$ is an indicator for the GCD of $f$ and $g$ being non-trivial. In this section, we will cover some basic and well-known facts about the resultant. The reader can refer to \cite{von_zur_Gathen_Gerhard_2013} for a more detailed account about resultants. The resultant of $f$ and $g$ is defined as follows.

\begin{definition}[Resultant]
    Let $f(Y), g(Y) \in \F[Y]$ be nonzero polynomials of
    degrees $d_f$ and $d_g$, respectively. Write
    \[
        f(Y) = \sum_{k=0}^{d_f} a_k Y^k,
        \qquad
        g(Y) = \sum_{k=0}^{d_g} b_k Y^k.
    \]
    The \emph{Sylvester matrix} of $f$ and $g$ is the
    $(d_f+d_g) \times (d_f+d_g)$ matrix $M$ 
    \[
    \begin{bmatrix}
    a_{d_f} & a_{d_f-1} & \cdots & a_0 & & \\
    & \ddots & & & \ddots & \\
    & & a_{d_f} & a_{d_f-1} & \cdots & a_0 \\
    b_{d_g} & b_{d_g-1} & \cdots & b_0 & & \\
    & \ddots & & & \ddots & \\
    & & b_{d_g} & b_{d_g-1} & \cdots & b_0
    \end{bmatrix}
    \]
The \emph{resultant} of $f$ and $g$, denoted by $\Res_Y(f,g)$, is defined as the determinant of the above matrix. 
\[
    \Res_Y(f,g) := \det(M).\qedhere
\]
\end{definition}

As mentioned previously, $\Res_Y(f,g)$ is zero exactly when the GCD of $f$ and $g$ is non-trivial. This fact is clearer from following equivalent characterization of the resultant.

\begin{lemma}[Poisson Formula]
    \label{lem:poisson-formula}
    Let $f(Y), g(Y) \in \F[Y]$ be nonzero polynomials of
    degrees $d_f$ and $d_g$ and suppose $\cB = \{\beta_1, \beta_2, \dots, \beta_{d_g}\}$ are the multiset of roots of $g$. Then,
    \[ \Res_Y(f, g) = (-1)^{d_f d_g} \operatorname{lc}(g)^{d_f}\prod_{i \in [d_g]} f(\beta_i)\]
    where $\operatorname{lc}(g)$ denotes the leading coefficient of $g$.
\end{lemma}

For bivariate polynomials $f, g \in \F[T, Y]$, $\Res_Y(f, g)$ is the resultant when $f$ and $g$ are viewed as polynomials in $\F[T][Y]$. In this case, $\Res_Y(f, g)$ would be a polynomial in $T$. From the definition of the resultant, we can bound the degree of $\Res_Y(f, g)$.

\begin{lemma}
    \label{lem:resultant-degree}
    Let $f, g \in \F[T, Y]$ be polynomials with $Y$-degrees $d_f$ and $d_g$ respectively. Suppose the $T$-degree of both the polynomials is at most $D$. Then, $\deg(\Res_Y(f, g))$ is at most $D(d_f + d_g)$.
\end{lemma}

\subsection{Deterministic Univariate Factorization}
\label{sec:univariate-factorization}

Over rational numbers, univariate factorization can be performed in deterministic polynomial time using the well-known LLL algorithm of Lenstra, Lenstra, and Lov\'asz~\cite{LLL82}. We define a convenient notation for the runtime of univariate factorization below, and then state the LLL theorem.

\begin{definition}
    \label{def:time-taken-to-factor-univariates}
    Let $f \in \F[X]$ be an univariate polynomial of degree $d$. We use $\cT(\F,d)$ to denote the time taken to factor $f$ over $\F$.
\end{definition}

\begin{theorem}[Univariate factorization over rationals {\cite{LLL82}}]
\label{thm:univariate-factorization-rationals}
There is a deterministic algorithm that, given a polynomial $f\in \Q[x]$ of degree $d$ and coefficient bit complexity at most $B$, computes the irreducible factorization of $f$ over $\Q$ in time $\poly(d,B)$. Thus, over $\Q$, $\cT(\Q,d)$ is polynomial in $d$ and in the bit complexity of coefficients of the input.
\end{theorem}

\subsection{Polynomial factorisation preliminaries}

The following is a standard pre-processing step that is performed in almost all factorisation algorithms (for instance, see \cite{BKRRSS25a} for more details). 

\begin{lemma}[Factorisation pre-processing]
    \label{lem:factorisation-preprocessing}
    Let $F(\X, Y) \in \F[\X,Y]$ be an $(n+1)$-variate polynomial with $\deg_Y F \leq d$ and $\deg F \leq D$, such that $F$ is computable by a circuit $C$ of size $s$ and depth $\Delta$. Then from the circuit $C$ for $F$, and access to a PIT oracle for $\poly(s,n,d)$-sized depth $\Delta + O(1)$ circuits, we can efficiently (in time $\poly(s,d,n)$) compute a circuit $\tilde{C}$ of size $\poly(s,n,d)$ and depth $\Delta + O(1)$ for a polynomial $\tilde{F}(\X, T, Y) \in \F[\X][T,Y]$ with the following properties:
    \begin{itemize}\itemsep 0pt
    \item $\tilde{F}$ is monic in $Y$ of $Y$-degree $d$ and total degree at most $D$,
    \item every monomial in $\tilde{F}$ that is divisible by any $\X$-variable is also divisible by $T$ and thus $\tilde{F}(\X, 0, Y) = \tilde{F}(\mathbf{0}, 0, Y)$,
    \item $\tilde{F}$ is square-free, and the polynomial $\tilde{F}(\mathbf{0}, 0, Y) = \tilde{F}(X, 0, Y)$ is also square-free.
    \end{itemize}

    Furthermore, the circuits for the factors of $F(\X,Y)$ can be efficiently recovered from circuits for the factors of $\tilde{F}(\X, T, Y)$. 
\end{lemma}

Thus, from this point on, our focus is to factorize a polynomial $f(T, Y) \in \F[\X][T, Y]$ (where $|\X| = n$) such that it is: (i) square-free (ii) monic in $Y$, with $Y$-degree at most $d$ (iii) $f(0, Y)$ is square-free (iv) computable by a circuit of size $\poly(s, n, d)$ and depth $\Delta$. We first state Gauss' lemma, which essentially states that it is equivalent to study the factorization of $f(T, Y)$ over $\F[\X][T, Y]$ and $\F(\X)[T, Y]$. The version we state below suffices for our setting, for a more general version refer \cite[Corollary 6.10]{von_zur_Gathen_Gerhard_2013}

\begin{lemma}[Gauss' lemma]
    Let $f \in \F[\X][T, Y]$ be monic in $Y$. Then $f$ is irreducible in $\F(\X)[T, Y]$ if and only if $f$ is irreducible in $\F[\X][T, Y]$.
\end{lemma}

Since $f(T, Y)$ is monic in $Y$, the irreducible factors of $f$ over $\F[\X]$ are also monic in $Y$. Hence, these irreducible factors are also irreducible over $\F(\X)$. Thus, the factorization of $f$ over $\F[\X]$ and $\F(\X)$ are the same. Hence, we will study the factorization $f(T, Y)$ over the field $\F(\X)$ (instead of the ring $\F[\X]$). Suppose $f(T, Y)$ factorizes as follows,
\[ f(T, Y) = f_1(T, Y) \cdot f_2(T, Y) \cdots f_k(T, Y) ,\]
where $f_1, \dots, f_k$ are irreducible polynomials in $\F(\X)[T, Y]$. Also, suppose $f(0, Y)$ factorizes as,
\[ f(0, Y) = (Y - \beta_1) \cdot (Y - \beta_2) \cdots (Y - \beta_d) ,\]
where $\beta_1, \dots, \beta_d \in \overline \F$. We recall the below standard lemma (known as Hensel's lemma in the literature) that $f(T, Y)$ completely factorizes into linear factors in power series roots in $Y$.

\begin{lemma}[Hensel's Lemma]
    \label{lem:hensel-lemma}
    Let $f(T, Y) \in \F[X][T, Y]$ be a polynomial which is monic in variable $Y$ such that $f(0, Y)$ is square-free. Suppose $f(0, Y)$ factorizes into,
    \[ f(0, Y) = (Y - \beta_1) \cdot (Y - \beta_2) \cdots (Y - \beta_d) ,\]
    where $\beta_1, \dots, \beta_d$ are distinct elements in $\overline \F$.
    Then, $f(T, Y)$ factorizes uniquely into,
    \[ f(T, Y) = (Y - \varphi_1(T)) \cdot (Y - \varphi_2(T)) \cdots (Y - \varphi_d(T)) ,\]
    where for each $i \in [d]$, $\varphi_i(T) \in \F(\beta_i)(\X) \indsquare{T}$ and $\varphi_i(0) = \beta_i$.
\end{lemma}

\cite[Theorem 4.1]{BKRRSS25a} showed how to obtain polynomial size and constant depth for the (truncations of) above power series roots $\varphi_i(T)$. We record this a theorem below for our setting, and note that this can be obtained by essentially invoking \cite[Theorem 4.1]{BKRRSS25a} for the shifted polynomial $f(T, Y + \beta_i)$.

\begin{theorem}[\cite{BKRRSS25a}]
    \label{thm:truncations-of-power-series-roots-of-constant-depth-circuits}
    Let $f(T, Y) \in \F[X][T, Y]$ be a polynomial as in the setting of \cref{lem:hensel-lemma}. Suppose $f$ is computable by a circuit of size $s$ and depth $\Delta$. Then for each $i \in [d]$,
    \[ \varphi_i(T) \bmod T^\ell ,\]
    is computable by a circuit $C$ of size $\poly(s, n, d, \ell)$ and depth $\Delta+O(1)$ over the field $\F(\X)(\beta_i)$. Moreover, there is an algorithm to compute $C$ with at most $\poly(s, n, d, \ell)$ arithmetic operations. 
\end{theorem}

The number of arithmetic operations specified above can be obtained by inspecting the proof of \cite[Theorem 4.1]{BKRRSS25a}. Using the above theorem, we can also obtain constant depth circuits for ``hensel lifts'' of the polynomial $f(T, Y)$. Suppose $f(0, Y)$ factorizes as,
\begin{align}
    f(0, Y) &= \tilde g_1(Y) \cdot \tilde g_2(Y) \cdots \tilde g_{r}(Y) \label{eqn:factorize-of-f-0-Y-into-irreducibles-1} \nonumber
\end{align}
where for each $i \in [r]$, $\tilde g_1, \tilde g_2, \dots, \tilde g_{r} \in \F[Y]$ are irreducible polynomials. The monic (in $Y$) Hensel-lift  from $\tilde g_1, \tilde g_2, \dots, \tilde g_{r} \in \F[Y]$ is the following factorization,
\[ f(T, Y) = g_1(T, Y) \cdots g_{r}(T, Y)\]
where $g_i(T,Y) \in \overline{\F}[\X]\indsquare{T}[Y]$ is given by 
\[ g_i(T, Y) = \prod_{j \in [d]: \tilde g_i(\beta_j) = 0} (Y - \varphi_j(T)).\]
Although each $\varphi_i(T)$ is a power series that may use coefficients from $\overline{\F}$, all coefficients of $g_i$ are symmetric functions in the roots of $\tilde{g}_i(Y)$ and thus $g_i(T,Y) \in \F[\X]\indsquare{T}[Y]$. 

Using \cref{thm:truncations-of-power-series-roots-of-constant-depth-circuits}, we can construct size $\poly(n,s,d,B)$ size and $\Delta+O(1)$ depth circuits $C_j$ for $\varphi_j(T) \bmod T^B$ over the field $\F(\beta_j)$. Using these circuits, we can build $\poly(n,s,d,B)$ size and depth $\Delta+O(1)$ circuits\footnote{Technically, we will get rational functions over $\{\beta_j : \tilde g_i(\beta_j) = 0\}$, where the $\beta_j$s can be thought of as formal variables, and both the numerator and the denominator have constant-depth circuits. But this can be handled to eventually get constant-depth division-free circuits for the Hensel lifts, based on the ideas discussed in \cite[Section 5]{BKRRSS25a}.} for $g_i(T, Y)$ over the field $\F(\beta_j : \tilde g_i(\beta_j) = 0)$. 
In order to obtain constant depth circuits for $g_i(T, Y)$ over the base field $\F$, we recall the following theorem from \cite{BKRRSS25b}.

\begin{theorem}[Complexity of symmetric polynomials \cite{BKRRSS25b}]
    \label{thm:complexity-of-symm-poly}
    Let $n,d,s \in \N$, and suppose that $\F$ is a field
    satisfying $|\F| \ge (nds)^c$ for an absolute constant $c > 0$.
    Suppose that $P \in \F[x_1,\ldots,x_n]$ is symmetric of degree $d$
    and that a circuit of size $s$ and depth $\Delta$ computes $P$.

    Write $\Esym_i(\X)$ for the $i$th elementary
    symmetric polynomial in $\X = (X_1,\ldots,X_n)$.
    Let $Q \in \F[Z_1,\ldots,Z_n]$ be the unique polynomial satisfying
    \[
        P(\X)
        = Q\bigl(\Esym_1(\X), \ldots,
                 \Esym_n(\X)\bigr).
    \]
    Then $\deg Q \le d$, and there is a circuit over $\F$ computing $Q$
    with size $\poly(s,d,n)$ and depth $\Delta + O(1)$.
\end{theorem}

\begin{remark}
    In fact, by inspecting the proof of \cref{thm:complexity-of-symm-poly} in \cite[Theorem 16]{BKRRSS25b}, given a size $s$ and depth $\Delta$ for $P$, the above theorem gives us an algorithm to compute the size $\poly(s,d,n)$ and depth $\Delta + O(1)$ circuit for $Q$ in $\poly(s,d,n)$ time.
\end{remark}

Suppose $a_{i, 1}, \dots, a_{i, l_i} \in \F$ are the coefficients of $\tilde g_i$. Observe that $g_i$ is symmetric over the roots of $\tilde g_i$ and the coefficients $a_{i, 1}, \dots, a_{i, l_i} \in \F$ are elementary symmetric polynomials (upto signs) in the roots of $\tilde{g}_i$. Thus, we invoke \cref{thm:complexity-of-symm-poly} to obtain a circuit of $\poly(n,s,d,B)$ size and $\Delta+O(1)$ depth computing $g_i$ which uses constants from $\F(a_{i, 1}, \dots, a_{i, l_i}) = \F$. From \cref{thm:truncations-of-power-series-roots-of-constant-depth-circuits} and \cref{thm:complexity-of-symm-poly}, all the above circuit constructions can be performed with at most $\poly(n,s,d,B)$ many arithmetic operations.

\begin{theorem}[Computing Hensel-lifts efficiently]
    \label{thm:hensel-lift-have-constant-depth-ckt}
    Let $f(T, Y) \in \F[X][T, Y]$ be a polynomial which is monic in variable $Y$ such that $f(0, Y)$ is square-free with the irreducible factorization $f(0,Y) = \tilde{g}_1(Y) \dots \tilde{g}_r(Y)$. Suppose $f$ is computable by a circuit of size $s$ and depth $\Delta$. Then, for any $B \geq 0$, we can compute (in time $\poly(s,n,d,B)$) circuits of size $\poly(s, d, n, B)$ and depth $\Delta + O(1)$ for polynomials $g_1(T,Y), \cdots, g_r(T,Y) \in \F[\X][T,Y]$, each of which are monic in $Y$, such that for every $i \in [r]$, $g_i(0,Y) = \tilde{g}_i(Y)$, and
    \[
    f(T, Y) = g_1(T,Y) \cdots g_r(T, Y) \bmod{T^B}.
    \]
\end{theorem}

\subsection{Standard algorithms in linear algebra}

We summarize some well-known and useful facts from Linear Algebra below. The reader can refer \cite[Sections~14.4--14.5]{Shoup2009} for a more thorough treatment of the topic.

\begin{lemma}
    \label{lem:find-kernel}
    Given a matrix $M\in\F^{m_1\times m_2}$ as input, there is an algorithm that computes a basis for $\ker(M)$ and performs atmost $\poly(m_1, m_2)$ field operations.
\end{lemma}
\begin{proofsketch}
    We can compute $\ker(M)$ by Gaussian Elimination.
\end{proofsketch}

\begin{lemma}
    \label{lem:intersection-of-subspaces} 
    Let $V,W \subseteq \F^m$ be two subspaces specified by their bases $\cB_1$ and $\cB_2$ respectively, then there is an algorithm to compute a basis $\cB$ for the space $V \cap W$. Moreover, the algorithm performs at most $\poly(m, |\cB_1|, |\cB_2|)$ many field operations.
\end{lemma}
\begin{proofsketch}
    Find a basis for the kernel of the matrix $[ \, A \,\,\, -B \,]$. Suppose $(x_1, y_1), \dots, (x_t, y_t)$ is the kernel of the matrix obtained from \cref{lem:find-kernel}. Then, $Ax_1, \dots, Ax_t$ is the required basis of $V \cap W$.
\end{proofsketch}

\begin{lemma}
    \label{lem:algo-for-0-1-basis}
    Let $V\subseteq\F^m$ be a vector space with basis $\cB$. 
    Suppose $V$ has a basis $\cB'$ such that $\cB'\subseteq\{0,1\}^m$ and every element in $\cB'$ has disjoint support. Then such a basis $\cB'$ is unique. 
    Moreover, given $\cB$ there is an algorithm which does at most $\poly(m)$ field operations and outputs $\cB'$.
\end{lemma}
\begin{proofsketch}
    Let $A, B$ be matrices whose rows are $\cB$ and $\cB'$ respectively.
    Suppose the rows in $B$ are ordered such that for rows $i, j$ (with $i < j$), the first $1$ in row $j$ does not appear before the first $1$ in row $i$. The lemma follows by noticing that $B$ is the reduced row echelon form (rref) of $A$ and rref of a matrix can be computed efficiently.
\end{proofsketch}

\section{Combining power-series into actual factors}\label{sec:logarithmic-derivative-method}

For this section, using \cref{lem:factorisation-preprocessing} if required, we may assume that $f(T, Y) \in \F[\X][T, Y]$ ($|\X| = n$) such that, (i) $\deg_Y(f) \leq d$ (ii) $\deg(f) \leq D$ (iii) $f$ is monic in $Y$ and square-free (iv) $f(0, Y) \in \F[Y]$ is square-free. 

Let $f(T, Y) = f_1(T, Y) \cdots f_k(T, Y)$ be the irreducible factorisation (over $\F$) of $f$ that we are trying to compute. Let the power-series $Y$-roots of $f$ be $\varphi_1(T), \ldots, \varphi_d(T) \in \overline{\F}[\X]\indsquare{T}$. Let the irreducible factorisation of $f(0,Y)$ over $\F$ be \[f(0,Y) = \tilde{g}_1(Y) \cdot \dots \cdot \tilde{g}_r(Y)\] which is also equal to $\prod_{i\in [d]}(Y-\varphi_i(0))$ over $\overline{\F}$. Using the $\tilde{g}_i(Y)$s, we compute the Hensel lifts (using \cref{thm:hensel-lift-have-constant-depth-ckt}) $g_1(T,Y), \dots, g_r(T,Y)$ satisfying

\[
f(T, Y) = g_1(T,Y) \cdots g_r(T,Y) \bmod T^B.
\]
where $B > 2D^2$. Each irreducible factor $f_i$ corresponds to a product of some block of $g_j$'s above. Thus, $[r] = S_1 \sqcup \cdots \sqcup S_k$ such that $f_i = \prod_{j\in S_i} g_j \bmod T^B$. Our goal is to try and find the blocks $S_1 \sqcup \cdots \sqcup S_k$. The following is a homogeneous linear system that would allow us to find these blocks. \\

\noindent
Consider the following homogeneous linear system over indeterminates $\mu_1, \cdots, \mu_r \in \F(\X)$
\begin{align*}
\mu_1 \cdot (\partial_Y g_1) \cdot g_2 \cdots g_r + \cdots + \mu_r \cdot g_1 \cdots g_{r-1} \cdot (\partial_Y g_r) = Q(T, Y) \bmod{T^B}
\end{align*}
for some polynomial $Q(T,Y)$ with $\deg_Y Q < \deg_Y f = d$ and $\deg Q \leq \deg f = D$. This is equivalent to ensuring that the coefficient of all $T^a Y^b$ of $\mu_1 \cdot (\partial_Y g_1) \cdot g_2 \cdots g_r + \cdots + \mu_r \cdot g_1 \cdots g_{r-1} \cdot (\partial_Y g_r)$ is zero whenever $a + b > D$ or $b \geq d$, which are homogeneous linear equations in the indeterminates $\mu_1,\cdots, \mu_r$.

\begin{theorem}[\cite{Sasaki1991, Sasaki1993, BLSSW04, Lecerf2006, Lec06, Lecerf2010}]
    \label{thm:kernel-charaterization-of-L-sigma} 
    Suppose $f(T, Y) \in \F[\X][T,Y]$ satisfies the conditions above and suppose $f(T, Y) = g_1(T, Y) \cdots g_r(T,Y) \bmod T^B$ with $B > 2 D^2$ where $\deg f \leq D$. 
    Let $\mathcal{L}_B$ denote the solution space to the above system $(\mu_1,\cdots, \mu_r) \in \F(\X)^r$. Then, 
    \[
    \mathcal{L}_B = \operatorname{span}\inbrace{\bm{1}_{S_1}, \dots, \bm{1}_{S_k}}.
    \]
\end{theorem}

We present a short proof here for completeness. 

\begin{proof}
We first prove the easier direction of showing that each $\bm{1}_{S_i}$ is a solution to the above system for $i \in [k]$. Indeed, since $\prod_{j \in S_i} g_j = f_i \bmod T^B$, we have
\begin{align*}
\sum_{j \in S_i} (\partial_Y g_j) \prod_{k\neq j} g_k = (\partial_Y f_i) \cdot \prod_{k \neq i} f_k \bmod T^B
\end{align*}
and the RHS is clearly a polynomial of $Y$-degree at most $d-1$ and total degree at most $D$. \\

\noindent
Suppose $\mu = (\mu_1, \cdots, \mu_r) \in \F(\X)^r$ be an arbitrary nonzero solution to the system
\[
\mu_1 \cdot (\partial_Y g_1) g_2 \cdots g_r + \cdots + \mu_r \cdot g_1 \cdots g_{r-1} (\partial_Y g_r) = Q(T, Y) \bmod{T^B}
\]
with $\deg_Y Q < d$ and $\deg Q \leq D$. We will show that $\mu$ is constant on coordinates in $S_i$. For simplicity, let us assume $1,2 \in S_1$ so that $g_1 g_2 \mid f_1$, and we shall show that $\mu_1 = \mu_2$.
Pick any root $\varphi_\ell(T)$ of $g_1(T, Y)$. Then, 
\begin{align*}
    Q(T, \varphi_\ell) & = \mu_1 \cdot (\partial_Y g_1)(T, \varphi_\ell) \cdot (g_2 \cdots g_r)(T, \varphi_\ell) \bmod{T^B}\\
    & = \mu_1 \cdot (\partial_Y f_1)(T, \varphi_\ell) \cdot (f_2 \cdots f_k)(T, \varphi_\ell)
\end{align*}
Therefore, for $G_1(T, Y) := Q(T, Y) - \mu_1 \cdot (\partial_Y f_1) f_2 \cdots f_k$, we have $G_1(T, \varphi_\ell) = 0 \bmod{T^B}$. Using Poisson Formula (\cref{lem:poisson-formula}), this implies that,
\begin{align*}
\Res_Y(G_1(T,Y), f_1(T, Y)) &= (-1)^{\deg_Y(G_1) \deg_Y(f_1)} \operatorname{lc}_Y(f_1)^{\deg_Y(G_1)} \prod_{j: f_1(T, \varphi_j) = 0} G_1(T, \varphi_j) = 0 \bmod{T^B}.
\end{align*}
On the other hand, by \cref{lem:resultant-degree}, $\Res_Y(G_1, f_1)$ is a polynomial in $T$ of degree at most $2D^2$. Hence, 
\[
\Res_Y(G_1(T,Y), f_1(T, Y)) = 0 \implies f_1(T, Y) \mid G_1(T,Y).
\]
Along exactly the same lines, we can also obtain that $G_2(T,Y) := Q(T,Y) - \mu_2 \cdot (\partial_Y f_1) f_2 \cdots f_k$ is divisible by $f_1$ as well. This implies that $f_1$ must divide $(\mu_1 - \mu_2) \cdot (\partial_Y f_1) f_2 \cdots f_k$ and this is possible only if $\mu_1 = \mu_2$. 
\end{proof}

In other words, $\mathcal{L}_B$ is precisely the space that is spanned by the indicator vectors corresponding to the blocks corresponding to irreducible factors of $f$. 

\subsection{Computing the kernel of the system using whitebox PIT}

In the previous section, we saw that the main \emph{recombination step} in factorization is reduced to computing the kernel of a certain matrix of polynomials. Moreover the kernel is known to be spanned by vectors in the base field.

In this section, we will see a general algorithm (by reduction to white-box PIT) for computing the $\F$-kernel of any matrix $M(\X) \in \F(\X)^{m_1 \times m_2}$ (where $|\X| = n$) whose entries consist of rational functions represented by some structured circuits. The $\F$-kernel of $M(\X)$, denoted by $\ker_{\F}(M(\X))$, is the subspace of all vectors $v \in \F^{m_2}$ such that $M(\X)v = 0$.

\begin{theorem}
\label{lem:compute-kernel-of-matrix-of-rational-functions}
Suppose we have an oracle for PIT for arbitrary linear combinations of entries of the matrix $M(\X) \in \F[\X]^{m_1 \times m_2}$. Then, the $\F$-kernel of $M(\X)$ can be computed with at most $\poly(n,d,m_1, m_2)$ many operations (including calls to the PIT oracle), where $d$ is an upper bound on the degree of the polynomial entries in $M$. 
\end{theorem}
\begin{proof}
The proof of the theorem is the following algorithm. Let us begin with $V = \F^{m_2}$ and we will keep pruning $V$ down to $\ker_{\F}(M(\X))$. In the general case, let $V \leq \F^{m_2}$ be spanned by $\inbrace{u_1, \cdots, u_\ell} \subseteq \F^{m_2}$, and we will always maintain the invariant that $\ker_{\F}(M(\X)) \subseteq V$. 

For each $i \in [\ell]$, we wish to check if $u_i \in \ker_{\F}(M(\X))$ or not. This amounts to checking if $M(\X)u_i = 0$ or not and this amounts to $m_1$-many PITs of $\F$-linear combinations of the entries of $M(\X)$. If one of these PITs is nonzero, we use the search-to-decision reduction for PIT (\cref{lem:decision-to-search-PIT}) to compute, in time $\poly(n,d)$, a point $\mathbf{a}$ such that $M(\mathbf{a})u_i \neq 0$, and prune $V$ to $V' = V \cap \ker(M(\mathbf{a}))$ (by using \cref{lem:find-kernel} and \cref{lem:intersection-of-subspaces}). Note that $V'$ continues to contain $\ker_{\F}(M(\X))$ but is now a smaller dimension subspace of $V$. However, if it happens that $M(\X)u_i = 0$ for all $i \in [\ell]$, then every $u_i$ is infact in $\ker_\F(M(\X))$ and thus we can safely return $\inbrace{u_1,\cdots, u_\ell}$ as a $\F$-basis for $\ker_\F(M(\X))$. 

This immediately yields a correct and efficient algorithm for computing $\ker_\F(M(\X))$, provided we have access to a PIT oracle for $\F$-linear combination of entries of $M(\X)$. 
\end{proof}

We summarise the above proof into the following algorithm. 

\begin{algorithm}[H]
\caption{Computing the $\F$-kernel of a matrix of arithmetic circuits}
 \label{alg:compute-kernel-of-matrix-relations}

\KwIn{$M(\X)\in\F[\X]^{m_1\times m_2}$ (where $|X| = n$), where each matrix entry is from a structured class of circuits. }

\KwOut{Basis for $\ker_\F(M(\X))$}

\KwOracle{PIT algorithm for $\F$-linear combinations of entries of $M(\X)$.}

\vspace{1em}

Initialise $V = \F^{m_2}$ given by basis $\mathcal{B} = \inbrace{e_1, \cdots, e_{m_2}} \subseteq \F^{m_2}$.\;

\For{$u_i \in \mathcal{B}$}{
    Run $m_1$ oracle calls (one for each row) to check if $M(\X)u_i = 0$. \;
    
    \If{$M(\X)u_i \neq 0$}{
        Use the PIT oracle to find a point $\mathbf{a} \in \F^{|\X|}$ such that $M(\mathbf{a})u_i \neq 0$. \;

        Replace $V$ with $V \cap \ker(M(\mathbf{a}))$, and $\mathcal{B}$ with a basis of $V$. Go to line 2. 
    }
}
\If{$M(\X)u_i = 0$ for all $u_i \in \mathcal{B}$}{
    \Return{the $\mathcal{B}$ as the basis for $\ker_\F{M(\X)}$.}
}
\end{algorithm}

\section{Factorization using White-box PIT Algorithm}
\label{sec:factorization-algorithm}

In this section, we will design the final factorization algorithm for constant-depth circuits which makes oracle calls to a white-box PIT algorithm. As we saw in the proof overview (\cref{sec:proof-overview}) and \cref{sec:logarithmic-derivative-method}, the logarithmic derivative method gives us a linear system, which specifies the appropriate truncated power series roots  to combine together to obtain the corresponding irreducible factors. 

\begin{theorem}
    \label{thm: Whitebox PIT implies Factorization} Let $n,D,s,\Delta\in\N$, $\F$ be the field of rational numbers. 
    
    There is an algorithm that takes as input an $n$-variate degree $D$ polynomial $F \in \F[\X]$ computed by a circuit of size $s$ and depth $\Delta$, requires oracle calls to a whitebox PIT algorithm for  $\poly(s, n, D)$-size and $\Delta+O(1)$-depth circuits, runs in $\poly(n,s,d,D)$ time, and outputs constant depth circuits for all the irreducible factors of $F$ together with their multiplicities. 

\end{theorem}
\begin{remark}
While stated and proved here for constant depth circuits, we note that \cref{thm: Whitebox PIT implies Factorization} also extends to other natural algebraic models like algebraic formulas, algebraic branching programs and algebraic circuits, with very minimal changes to the proof. 
\end{remark}
\begin{remark}
    One can also keep track of the bit complexity of the rational numbers that show up in the computation in the following proof. Since all our computations involve ``constant-depth computations'' over input coefficients, the bit-complexities for any numbers generated in the algorithm will remain polynomially bounded in the input parameters. We skip this bookkeeping for clarity of exposition.
\end{remark}

\begin{proof}[Proof of \cref{thm: Whitebox PIT implies Factorization}]
    For ease of exposition, we will work with an $(n+1)$-variate polynomial $F(\X, Y)$, where the last variable is specially denoted by $Y$. Suppose $F$ is computable by a circuit of size $s$ and depth $\Delta$. By the discussion in     \cref{lem:factorisation-preprocessing}, 
    we can reduce factoring $F(\X, Y)$ to factoring a polynomial $f(T, Y) \in \F[\X][T, Y]$ ($|\X| = n$) such that, (i) $\deg_Y(f) \leq d$ (ii) $\deg(f) \leq D$ (iii) $f$ is monic in $Y$ and square-free (iv) $f(0, Y) \in \F[Y]$ is square-free (v) $f(T, Y)$ is computable by a circuit of size $\poly(s, n, D)$ and depth $\Delta+O(1)$. So henceforth, we will aim to factor $f(T, Y)$.

    Suppose $f(T, Y)$ factorizes as follows,
    \begin{align}
        f(T, Y) &= f_1(T, Y) \cdot f_2(T, Y) \cdots f_k(T, Y) \label{eqn:factorize-into-irreducible-polynomials-1} .
    \end{align}
    where (i) for all $i \in [k]$, $f_i$ is an irreducible polynomial in $\F(X)[T, Y]$.
    And, suppose $f(0, Y)$ factorizes as,
    \begin{align}
        f(0, Y) &= \tilde g_1(Y) \cdot \tilde g_2(Y) \cdots \tilde g_{r}(Y) \nonumber
    \end{align}
    where for each $i \in [r]$, $\tilde g_1, \tilde g_2, \dots, \tilde g_{r} \in \F[Y]$ are irreducible polynomials.

    We can obtain this factorization of $f(0, Y)$ in $\poly(n,s,D)$ time as follows: Obtain coefficients of $f(0, Y)$ from $C$ using the interpolation lemma (\cref{lem:interpolation}) in $\poly(n, s, D)$ time. Then from the coefficients of $f(0, Y)$, we invoke the efficient deterministic univariate factorization algorithm over rationals in \cite{LLL82} to factor $f(0,Y)$ over $\F$ (\cref{thm:univariate-factorization-rationals}).

    Then, we Hensel lift from $\tilde g_1, \tilde g_2, \dots, \tilde g_{r} \in \F[Y]$ to obtain the following factorization,
    \[ f(T, Y) = g_1(T, Y) \cdots g_{r}(T, Y) \bmod T^{B}\]
    where $B = 2D^2+1$ and $g_i(T, Y) \in \F(X)[T, Y]$. 
    By \cref{thm:hensel-lift-have-constant-depth-ckt}, we can build circuits of size $\poly(s,n,D)$ and depth $\Delta+O(1)$ for $g_i(T, Y)$ over the base field. Moreover, this construction requires at most $\poly(s,n,D)$ arithmetic operations.

    As observed in \cref{sec:logarithmic-derivative-method}, we would like to find a partition $[r] = S_1 \sqcup \cdots \sqcup S_k$ such that $$f_i := \left(\prod_{j\in S_i} g_j\right) \bmod T^{D+1} .$$ By \cref{thm:kernel-charaterization-of-L-sigma}, using the circuits for $g_i(T, Y)$, we can build a matrix $M \in \F[\X]^{\poly(d,D) \times \poly(d,D)}$ whose $\F$-kernel is $\operatorname{span}_\F\inbrace{\bm{1}_{S_1}, \dots, \bm{1}_{S_k}}$. From the description of the linear system in \cref{thm:kernel-charaterization-of-L-sigma}, it is clear that this matrix can be constructed with at most $\poly(n,s,D)$ many arithmetic operations.

    By \cref{lem:compute-kernel-of-matrix-of-rational-functions}, we can find a basis $\cB$ for the $\F$-kernel of $M$ with at most $\poly(n,s,D)$ many arithmetic operations (with oracle calls to white-box PIT algorithms for $\poly(s,d,n)$ size and $\Delta+O(1)$ circuits). By \cref{lem:algo-for-0-1-basis}, we can obtain $\bm{1}_{S_1}, \dots, \bm{1}_{S_k}$ from the $\cB$ in $\poly(d)$ operations. Thus, we can obtain circuits of size $\poly(n,s,D)$ and depth $\Delta+O(1)$ for 
    \[ f_i := \left( \prod_{j\in S_i} g_j \right) \bmod T^{D+1} \]
    from the circuits of $g_j$, and clearly this can be done with at most $\poly(n,s,D)$ many arithmetic operations.

    To recover the multiplicity of an irreducible factor $f_i$, we can perform divisibility tests to compute the highest power of $f_i$  which divides $f$. Previous works (see \cite{Forbes15,BKRRSS25a}) discuss the reduction of divisibility testing of constant-depth circuits to PIT for constant-depth circuits, and it is easy to observe that this reduction also works in the whitebox setting.
    
This completes the proof of the theorem. 
\end{proof}

\bibliographystyle{alpha}
\bibliography{ref.bib}
\end{document}